# End-to-End QoS Improvement of HSDPA End-User Multi-flow Traffic Using RAN Buffer Management

Suleiman Y. Yerima and Khalid Al-Begain
Integrated Communications Research Centre, University of Glamorgan
Pontypridd (Cardiff) CF37 1DL, Wales, UK
E-mail: {syerima,kbegain}@glam.ac.uk

*Abstract*— **High Speed Downlink Packet Access (HSDPA) was introduced to UMTS radio access segment to provide higher capacity for new packet switched services. As a result, packet switched sessions with multiple diverse traffic flows such as concurrent voice and data, or video and data being transmitted to the same user are a likely commonplace cellular packet data scenario. In HSDPA, Radio Access Network (RAN) buffer management schemes are essential to support the end-to-end QoS of such sessions. Hence in this paper we present the end-to-end performance study of a proposed RAN buffer management scheme for multi-flow sessions via dynamic system-level HSDPA simulations. The scheme is an enhancement of a Time-Space Priority (TSP) queuing strategy applied to the Node B MAC-hs buffer allocated to an end user with concurrent real-time (RT) and non-real-time (NRT) flows during a multi-flow session. The experimental multi-flow scenario is a packet voice call with concurrent TCP-based file download to the same user. Results show that with the proposed enhancements to the TSP-based RAN buffer management, end-to-end QoS performance gains accrue to the NRT flow without compromising RT flow QoS of the same end user session.**

*Keywords-HSDPA; Buffer Management; End-toEnd QoS; Multi-flow Traffic; Performance*

## I. INTRODUCTION

In the past few years, UMTS cellular networks have been deployed on a large scale. UMTS was designed to support a variety of services with peak data rates of up to 2 Mb/s. In 2005, technical improvements to UMTS radio access network (UTRAN) downlink, collectively termed High Speed High Speed Downlink Packet Access (HSDPA) were introduced to support growing demand for broadband services. HSDPA provides peak data rates of up to 14 Mb/s, lower transmission latency, improved Quality of Service, and enhanced spectral efficiency for UMTS downlink traffic [1], [2],[3].

HSDPA was initially envisaged to support asymmetric data services such as internet browsing. But with the growing popularity of packet voice and multimedia services, coupled with the availability of advanced receivers, sessions with multiple flows of for example voice, video, and data to a single end user is a possible scenario on HSDPA. 3GPP documents [4] allow for the provision of separate data buffers in the Node B MAC-hs entity which is responsible for packet scheduling and also error recovery via a Hybrid Automatic Repeat Request (H-ARQ) mechanism.

The necessity to queue data packets in the Node B MAC-hs for packet scheduling and HARQ provides opportunity to apply buffer management schemes to improve end-to-end Quality of Service for the end users. This is particularly important for intra-user multi-flow sessions because the diverse flows usually have different QoS requirements. Furthermore, incorporating buffer management strategy in the Node B is likely to have significant impact on traffic and system performance because of its location at the edge of the network bottleneck.

Taking the aforementioned into account, in our previous work [5] we proposed an Intra-user MAC-hs buffer management scheme based on a Time-Space Priority (TSP) queuing mechanism as a solution for joint intra-user multi-flow QoS optimization. By comparative analysis we showed the potential of the TSP scheme for effective multi-flow QoS control. In this paper we present new enhancements to the original TSP scheme in [5] and focus on the resulting end-to-end QoS performance improvement. We discuss performance results from dynamic end-to-end system-level HSDPA simulations illustrating the QoS improvement achievable with the enhanced scheme, especially to TCP-based NRT traffic in the intra-user multi-flow session.

The rest of the paper is organized as follows. Section II describes HSDPA, Section III explains TSP buffer management and the proposed enhancements. Simulation of HSDPA scenarios for end-to-end performance analysis feature in Section IV, while results are discussed in Section V. Lastly, in Section VI we present the conclusions.

## II. BASIC FEATURES OF HSDPA

HSDPA utilizes a shared channel (HS-DSCH) to transmit data to the User Equipments (UE) over the downlink of a HSDPA enhanced UMTS cell. A HSDPA network consists of three interacting domains; Core Network (CN), UMTS Terrestrial Radio Access Network (UTRAN) and the UE i.e. the receiver. The Core Network is responsible for switching, transit and routing of user traffic. UTRAN provides the air interface access for the receiver and handles all radio related functionalities. UTRAN consists of a Radio Network Controller (RNC) and base station or Node B. The main features of HSDPA include Adaptive Modulation and Coding (AMC), Hybrid-ARQ, and Packet Scheduling which are all Node B functionalities.

AMC changes the modulation and coding scheme for data transmission in accordance with the variations in channel con-

This work is supported in part by the UK Overseas Research Award Scheme (ORSAS).

ditions of the UE. The *Transport Block* is the basic unit of data transmission governed by the AMC. The *Transport Block Size* (TBS) varies with the selected AMC scheme. Thus, when a HSDPA receiver has good channel conditions, AMC allows for transmission of a larger TBS. Channel conditions are estimated by the receiver using a *Channel Quality Indicator* (CQI) feedback via the uplink control channel i.e. High Speed Dedicated Physical Control Channel (HS-DPCCH). AMC scheme is based on CQI which indicates the maximum transport block size that can be received correctly with at least 90% probability. The HARQ functionality is responsible for retransmission of data that the receiver is unable to decode due to transmission errors. A 2ms transmission time interval (TTI) allows for fast scheduling of packet transmission to multiple users over the shared radio channel, fast tracking of the users' channel quality for the AMC and fast re-transmission for HARQ.

### III. INTRA-USER RAN BUFFER MANAGEMENT

3GPP HSDPA specifications do not include buffer or queue management algorithms for intra-user multiflow traffic leaving it as an open implementation-specific issue. As mentioned earlier, Node B buffer management is crucial to end-to-end QoS provisioning of intra-user multiple-flow sessions. However, the 3GPP specifications include mechanisms that can be exploited to support advanced buffer management solutions. For example, support for intra-user service differentiation with separate data queues in the RLC and MAC-hs layers [6]. Taking this into account, we designed a time-space priority (TSP) buffer management scheme in our earlier work [5] and showed its potential for joint QoS control compared to other possible HSDPA Node B buffer management schemes for intra-user RT and NRT multi-flow sessions. In this section, we describe TSP and our proposed new enhancements that further exploit existing 3GPP standards to improve end-to-end NRT throughput.

#### A. The Time Space Priority(TSP) Scheme

The Time-Space priority buffer management is a priority queuing scheme designed for joint QoS control of concurrent RT and NRT flows. Unlike most priority queuing that provide either delay or loss differentiation, the core concept of TSP is combined loss and delay differentiation in a single queue, thus yielding transmission (time) priority for RT packets and space priority for NRT packets [7]. A threshold ***R*** is used to partition the queue as shown in Fig. 2, such that NRT packets are accorded more buffer space by default since they are loss sensitive but delay tolerant. Note that ***R*** is not a hard partition but a threshold that limits the total number of RT packets admitted into the queue at any given time to ***R***. Thus, the maximum number of admitted NRT packets at any given time can vary from $N - R$ to $N$, where $N$ is the total queue capacity. This allows for more efficient buffer space utilization and further minimization of NRT packet loss.

Also, as shown in Fig. 2, TSP allows RT packets, such as video or voice packets, to be queued in front of the NRT packets for transmission priority because of their stringent delay requirements, and, since this ensures lower RT queuing delay, RT packets will not require as much buffering as NRT packets hence their lower space priority relative to NRT packets.

Due to loss sensitivity, arrival of NRT packets at the RNC necessitates the use of RLC Acknowledged Mode (AM) for onward transmission over the Iub interface to the Node B. RLC AM packets require feedback from peer RLC entity in the UE which is typically sent via a STATUS message [8]. NRT packets lost due to Node B buffer overflow can be recovered with the RNC-based RLC (selective repeat) ARQ retransmissions. Since the ARQ mechanisms operate between the RNC and recipient UE as depicted in Fig. 1, retransmissions increase the RLC round-trip-time resulting in overall end-to-end delay of NRT packets, which manifests in severe degradation of end-to-end-throughput for TCP-based NRT flow within a multi-flow session. This is undesirable since majority of NRT traffic utilize TCP as transport protocol. Furthermore, retransmissions lead to waste of Iub resources, Node B buffer space as well as air interface bandwidth. Hence, despite the space priority accorded to NRT packets in the TSP scheme, the aforementioned problem provide incentive for further enhancement in order to optimize performance in HSDPA multi-flow session scenario. Next we describe a TSP enhancing solution which exploits Iub flow control signaling. The end-to-end performance is compared to that of the original TSP scheme in section IV.

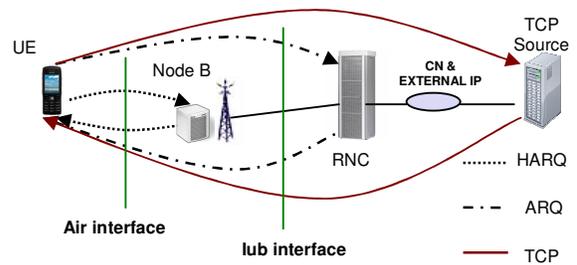

Figure 1. Packet Retransmission mechanisms in HSDPA. AM packet Losses due to Node B buffer overflow will require RLC level ARQ retransmissions leading to waste of bandwidth, buffer space and increased round-trip-time.

#### B. Enhancements to the TSP Scheme

We enhance the TSP scheme as shown in Fig. 2, with the addition of two thresholds ***L*** and ***H*** and a credit allocation algorithm designed to enable more efficient utilization of buffer space and air interface resources with minimal Iub signaling load. The allocation algorithm utilizes the NBAP signaling [9] as shown in the bottom part of Fig. 2, to issue credits to the RNC that determine the number of arriving packet data units (PDUs) to the Node B for each flow in the user's session. The number of credits per transmission interval (TTI) are determined as follows. Let the total credits per TTI for a particular multi-flow user be given by $C_{Total} = C_{NRT} + C_{RT}$ where:

$$C_{RT} = (\lambda_{RT} / PDU\_size) \cdot TTI \quad (1)$$

$C_{RT}$ is the number of credits per TTI for the RT flow in the multi-flow session, while PDU_size denotes the size of the PDU in bits. $\lambda_{RT}$ is the minimum guaranteed bit rate of the RT flow ( in bits/s), a parameter which can be obtained from the QoS attributes of the flow during bearer negotiation [6]. For the NRT flow, $C_{NRT}$ is given by:

$$C_{NRT} = \min \{ C_{NRTmax} , UBS_{NRT} \} \quad (2)$$

Where $UBS_{NRT}$ is the user's NRT buffer occupancy in the RNC and $C_{NRTmax}$ is the maximum NRT grants per TTI which depends on the HSDPA channel load, scheduling policy, and the recipient UE radio conditions. $C_{NRTmax}$ is calculated from:

$$C_{NRTmax} = (\lambda'_{NRT}/PDU\_size) \cdot TTI, \quad N_T < L$$

$$k \cdot (\lambda'_{NRT}/PDU\_size) \cdot TTI, \quad L \leq N_T \leq H$$

$$0, \quad N_T > H \quad (3)$$

Where **L** and **H** are the additional control thresholds in Fig. 2 and $N_T$ is the total number of RT and NRT PDUs in the queue. $k \in \{0,1\}$ is a factor for overflow control. $\lambda'_{NRT}$ is an estimate of the user's NRT data rate allocated by the packet scheduler in the MAC-hs. The estimate is obtained by using an exponentially weighted moving average filter according to:

$$\lambda'_{NRT} = \alpha \cdot \lambda'_{NRT-1} + (1-\alpha) \cdot \lambda_{NRT} \quad (4)$$

$\lambda_{NRT}$ is the instantaneous NRT bit rate. With (4), NRT grant allocation is made dependent on load and user channel quality which is appropriate because of the elastic nature of the NRT flow. Since averages are used in the grant calculation, the space between **H** and **N** absorbs instantaneous burst arrivals.

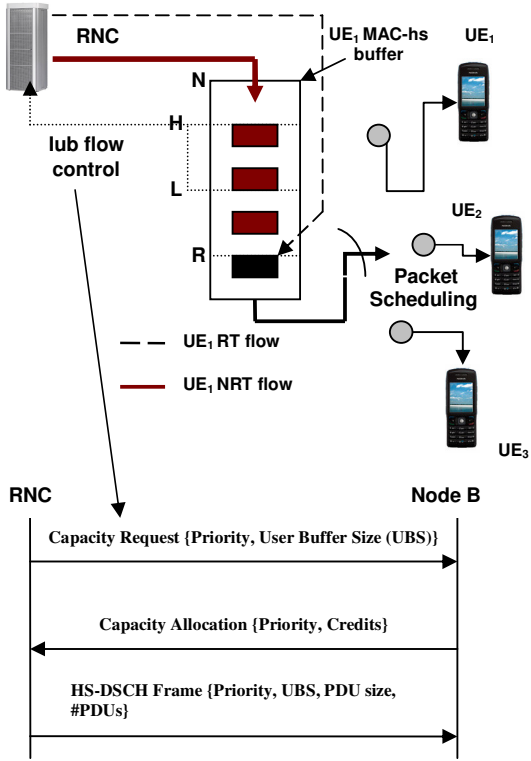

Figure 2. HSDPA UTRAN model with Time-Space Priority buffer management scheme enhanced with credit-based Iub flow control and a new credit allocation algorithm.

## IV. END-TO-END HSDPA SIMULATION

In order to evaluate the end-to-end performance improvement of the enhanced TSP scheme, we developed a custom system-level end-to-end HSDPA simulation model using OPNET modeler. We chose simulation as a modeling tool in order to represent as much detail as possible in the model and capture the dynamics of the scenarios more realistically. The simulation model is shown in Fig. 3.

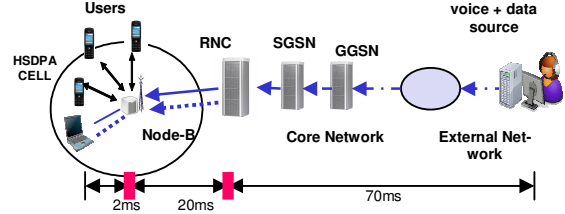

Figure 3. End-to-end HSDPA simulation set up

We simulated several scenarios where an end-user under observation in a HSDPA cell, is assumed to be downloading a file whilst in a voice conversation. Thus to model the multi-flow session, we implemented a VoIP packet source and an FTP source with TCP Reno. The VoIP is modeled as an ON/OFF source with same parameters from [10].

Other aspects of HSDPA modeled in detail include: RNC, with packet segmentation, RLC MAC queues, RLC AM and UM modes including ARQ for AM mode. RNC – Node-B Iub signaling is also modeled. In the Node-B, MAC-hs queues (applying TSP and enhanced TSP schemes), HARQ processes, AMC schemes, and Packet Scheduling on the HSDPA air interface are modeled. In the receiver, we included SINR calculation and CQI reporting, HARQ processes, RLC modes with ARQ for AM, packet reassembly queues, peer TCP entity, and an application layer.

In the experiments a test user equipment was connected to the HSDPA UTRAN through which multi-flow traffic was received in a simultaneous 120s voice conversation and file download session. VoIP packets were being received while file download was taking place using FTP over TCP. The overall set up models a single HSDPA cell. Radio link simulation included path loss and shadowing models with transmit powers and AMC schemes setting as given in Table 1. Number of available H-SDSCH codes is assumed to be 5, while CQI feedback latency was set to 6ms. Four HARQ processes were used in the HARQ manager, while Round Robin scheduling was employed in the packet scheduler. The performance metrics observed include:

- *End-to-end NRT throughput*: the end-to-end file download TCP throughput at the test UE receiver during the concurrent VoIP and file download multi-flow session.

- *End-to-end VoIP delay*: The end-to-end delay of VoIP packets measured in the multi-flow test UE receiver during the concurrent VoIP and file download session.

TABLE I. SUMMARY OF ASSUMED SIMULATION PARAMETERS.

| HSDPA Simulation Parameters | |
|---|---|
| HS-DSCH TTI | 2ms |
| Path loss Model | 148 + 40 log (R) dB |
| Transmit powers | Total Node B power=15W, HSDSCH power= 50% |
| Shadow fading | Log-normal: σ = 8 dB |
| AMC schemes | QPSK ¼, QPSK ½, QPSK ¾, 16QAM ¼, 16 QAM ½ |
| Number of assigned HSDSCH codes | 5 |
| CQI delay | 3 TTIs (6ms) |
| HARQ processes | 4 |
| HARQ feedback delay | 5ms |
| Test UE position from Node B | 0.2 km |
| Packet Scheduling | Round Robin |
| MAC PDU size | 320 bits |
| Iub (RNC-Node B) delay | 20ms |
| External + CN delays | 70ms |
| HS-DSCH frame | 10ms |
| Buffer Mgt. parameters | TSP: R= 10; N = 150 PDUs<br>E-TSP: R= 10; L=30; H=100; N=150 PDUs |
| Flow control parameters | α = 0.7; k = 0.5 |
| TCP Parameters: | Reno; MSS = 536 bytes; RWIND = 64 |

## V. RESULTS AND DISCUSSIONS

### A. Multi-flow End-user NRT End-to-End Performance

Figs. 4 to 6 show results of the end-to-end TCP layer throughput measurements for a test receiver during a 120s multi-flow session for various HSDPA cell loads. The test UE is assumed to be located 200m away from the base station, while other UEs are placed at random positions in the cell. Other users are assumed to be receiving a single flow of FTP downloads during their sessions and hence no buffer management scheme is applied to their MAC-hs queues.

Fig 4. plots the NRT throughput obtained without application of buffer management for the test multi-flow user in the MAC-hs; instead, arriving NRT or RT PDUs from the RNC for the test UE are queued and transmitted in a FIFO manner. We observe the drop in the test UE throughput when additional users are scheduled on the HSDPA channel. Since Round Robin scheduling is used, the throughput of the multi-flow UE is expected to drop with more users as the end-to-end TCP RTT increases due to increased inter-scheduling gaps, loss recovery at the RLC layer when MAC-hs buffer overflow occurs and also loss recovery at the TCP layer in the event that RLC layer recovery fails (after a maximum of six attempts).

The same experiment is repeated with TSP applied to the MAC-hs buffer of the test UE and the results are shown in Fig. 5. A similar pattern to Fig. 4 is observed and the same reasons apply for the observed behavior. Although, in Fig. 5 it can be observed that end-to-end RTT (and hence throughput) variation is lower with the TSP scheme compared to Fig. 4.

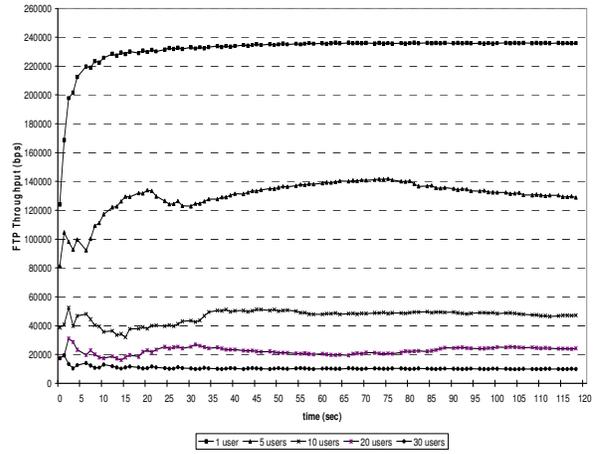

Figure 4. End-to-end throughput at test UE for a multi-flow session without buffer management over the session period. . Also shown are test UE throughput when 5, 10, 20 and 30 users share the HSDPA channel.

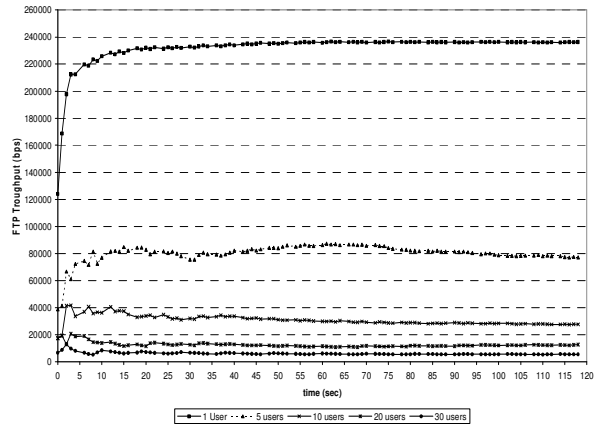

Figure 5. End-to-end throughput at test UE for multi-flow session with TSP buffer management over the session period. . Also shown are test UE throughput when 5, 10, 20 and 30 users share the HSDPA channel.

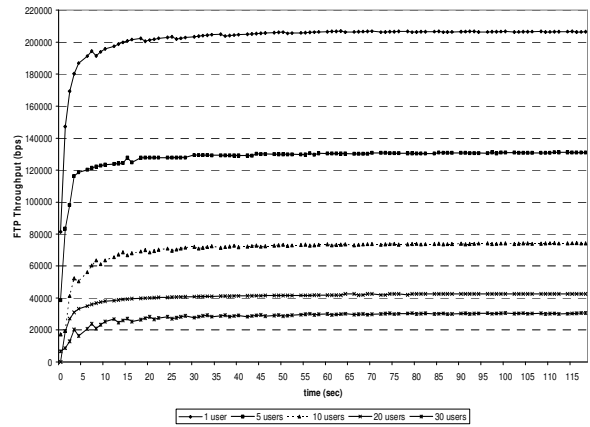

Figure 6. End-to-end Throughput at test UE for multi-flow session with enhanced TSP buffer management over the session period. Also shown are test UE throughput when 5, 10, 20 and 30 users share the HSDPA channel.

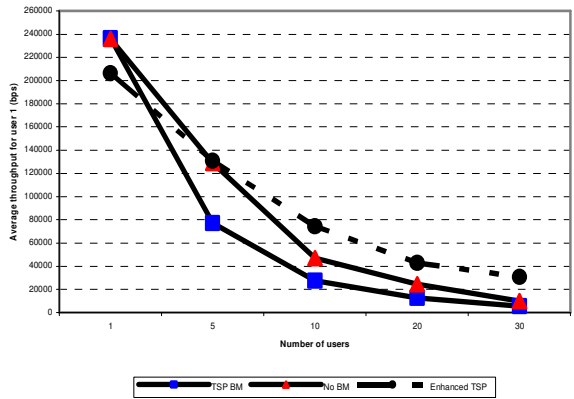

Figure 7. Average end-to-end Throughput at test UE for Vs number of users in cell. Results given for no BM, TSP BM and enhanced TSP BM.

In fig. 6 results of the same experiment with the enhanced TSP scheme applied to the MAC-hs queue of the multi-flow user is shown. The multi-flow user TCP throughput is seen to have lower throughput variation, indicating comparatively lower RTT variation. This is because the enhanced TSP was able to mitigate MAC-hs NRT PDU losses, reducing the likelihood of RLC level and hence TCP retransmissions. In Fig. 7 we illustrate the end-to-end throughput observed in the test multi-flow UE averaged over the entire session, for all the buffer management scenarios in a single graph. It shows that as the cell load (i.e. number of users) increases, the enhanced TSP scheme yields significant throughput performance improvement over the TSP scheme and the scenario with no buffer management. It is important to also note the corresponding lower end-to-end delay of the VoIP flow with the enhanced TSP compared to the one without multi-flow buffer management from Fig. 8, since we are considering concurrent RT and NRT user multi-flow session where the presence of one flow is expected to have an effect on the QoS of the other.

It is interesting to observe from Fig. 7 that the enhanced TSP gave the lowest throughput for the single user scenario. This reason for this is the increased TCP RTT as a result of the flow control mechanisms and the time priority of the RT PDUs. But as the cell load increases, the effect of the inter-scheduling gaps, losses due to MAC-hs buffer over flow and the resulting RLC retransmissions become more pronounced. Since the enhanced TSP is best able to cope with this, the performance improvement is noticeable at higher load.

At lower load the FIFO scheduling without buffer management is able to have a higher throughput compared to the original TSP, but cannot guarantee RT QoS at the same time. This is because NRT packets have fairer chance of scheduling opportunity at the expense of VoIP delay as depicted in Fig. 8. On the other hand, the trade-off in NRT throughput as a result of TSP schemes manifests in better VoIP delay performance.

### B. Multi-flow End-user RT end-to-end performance

The most important observation to be made from fig. 8 is that the end-to-end VoIP QoS is not degraded as a result of the enhancements to the TSP scheme. Whereas, we observe from Fig. 7 the corresponding end-to-end NRT QoS improvement resulting from the enhancement. While TSP and enhanced TSP show identical VoIP performance in the multi-flow session because of the time priority mechanism, they also maintain a fairly low variation in end-to-end delay with increasing load, compared to the scenario without buffer management. This underscores the need for buffer management solutions for joint QoS optimization during end-user multi-flow sessions, especially under higher load conditions.

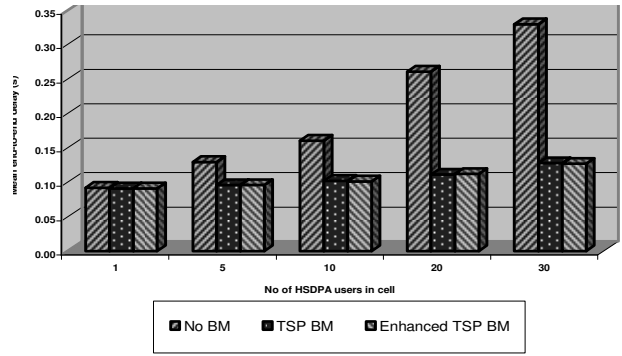

Figure 8. Average end-to-end VoIP PDU delay at test UE Vs number of users in cell. Results given for No BM, TSP and enhanced TSP.

### VI. CONCLUDING REMARKS

This paper presents enhancements to a Time-Space-Priority based buffer management scheme for intra-user multi-flow traffic control in HSDPA. By means of VoIP and TCP traffic simulation on an end-to-end HSDPA model developed in OPNET we showed that the enhancements can yield significant improvement in the end-to-end NRT performance without compromising the achievable end-to-end RT QoS.